\documentclass[aps,showpacs,preprintnumbers,amsmath, amssymb]{revtex4}

\usepackage{float}
\usepackage{graphics,epsfig}
\usepackage{graphicx}
\usepackage{dcolumn}
\usepackage{bm}
\usepackage{slashed}

\begin{document}
\baselineskip=0.8 cm

\title{{\bf Observing the inhomogeneity in the holographic models of superconductors}}
\author{Xiao-Mei Kuang$^{a}$}
\email{xmeikuang@gmail.com}
\author{Bin Wang$^{a}$}
\email{wang_b@sjtu.edu.cn}
\author{Xian-Hui Ge$^{b}$}
\email{gexh@shu.edu.cn}
\affiliation{$ ^{a}$CAA and Department of Physics and Astronomy, Shanghai Jiao Tong University, Shanghai 200240, China\\
$ ^{b}$Department of Physics, Shanghai University, Shanghai 200444, China}

\vspace*{0.2cm}
\begin{abstract}
\baselineskip=0.6 cm
\begin{center}
{\bf Abstract}
\end{center}

We study the gravity duals of striped holographic superconductors in the AdS black hole and AdS soliton backgrounds.  We show the dependences of the condensation and the critical temperature/critical chemical potential on the inhomogeneity in these two different spacetimes. By exploring the dynamics of the normal phase through the scalar field perturbation, we argue that the pair susceptibility and the conductivity can be possible phenomenological indications to disclose the property of inhomogeneity.

\end{abstract}

\maketitle
\newpage
\vspace*{0.2cm}

\section{Introduction}
High temperature superconductor is one of the most exciting scientific discoveries in the past thirty years.  However the theoretical understanding of it is still lacking. The difficulty arises from the fact that these high temperature superconductors are strongly coupled. Recently there has been a flurry of attempts trying to understand such strongly coupled superconductors by using various developments of the gauge/gravity correspondence (for reviews on this topic, see for examples \cite{Maldacena,S.S.Gubser-1,E.Witten} and references therein). In the gauge/gravity duality, the strongly coupled condensed matter systems are mapped to a weakly coupled gravitational theory in the AdS spacetimes.  A particular model \cite{3Hprl}  consists of a 3+1-dimensional Einstein-Maxwell-scalar theory in an AdS black hole background and the superconducting phase transition can be formed when scalar hairs condensate on the black holes at low temperatures.  This model is called holographic superconductor which can be used to deal with the strongly coupled system.

The available holographic superconductor models cannot account for the effect of inhomogeneity in  strongly coupled system.  From the numerical calculations of the Hubbard models \cite{hubbard1}, it was observed that inhomogeneity plays a much bigger role in strongly coupled superconductors than in standard, weakly coupled superconductors. Recently there have been a lot of experimental evidences showing that the inhomogeneous superconductors exhibit a lot of peculiar properties compared with the homogeneous superconductors \cite{ConStriped1,ConStriped2,ConStriped3,ConStriped4}.  Thus in addition to the strong coupling, the inhomogeneity is also important to show the path to the mystery of high temperature superconductors.

It is of great interest to devise a comprehensive holographic model to incorporate the strong coupling and the inhomogeneity in decribing the high temperature superconductors.  Some progresses have been made along this direction \cite{Pajer,Rozali,Siopsis1,Siopsis2,Donos} in taking the holographic superconductor model one step closer to real systems by considering inhomogeneous configurations. Studying the inhomogeneous solutions of the holographic superconductor with  a modulated chemical potential of wave vector $Q$, in the probe limit it was shown that below a critical temperature superconducting stripes develop and the critical temperature depends on the modulation's wave vector, which characterizes the inhomogeneity \cite{Pajer,Rozali,Siopsis1}. The effects of the inhomogeneity on the superconducting transition temperature by including the effects of backreaction onto the spacetime geometry was also examined in \cite{Siopsis2}.

Besides the condensation and the critical temperature, we would like to ask whether there is an effective phenomenological way to describe the inhomogeneity in the superconducting phase transitions. In \cite{kuang-prd} it was argued that the conductivity and the pair susceptibility are  two possible probes to distinguish the order of phase transition in the holographic model of superfluidity. Since these two quantities are measurable in condensed matter physics, it was found that they can help us understand more of the phase structure in the holographic superfluidity.  In this work, we are going to examine whether these two phenomenological probes are useful to disclose signatures of inhomogeneities in the holographic superconductor.  We will show that in addition to the conductivity, the pair susceptibility is a possible tool to uncover  the inhomogeneity.  These phenomenological quantities can help us understand more of the property of inhomogeneity in the holographic model of superconductor.

Furthermore,  in this work we will not restrict our discussions in the black hole background. We would like to extend the discussion of the inhomogeneity to the  AdS soliton background. First attempt in this direction was reported in \cite{Ge}.  We will  show more behaviors of the strongly coupled striped insulator and its dependence on the inhomogeneity. Moreover we will examine  whether there are phenomenological signatures to show the imprints of inhomogeneity in the AdS soliton background. Meanwhile, we  study the properties of the pair susceptibility and the conductivity in this system.  As we discuss in Section IV, we find they are very different from the results in the black hole. Early investigation of the pair susceptibility about inhomogeneous system in condensed matter can be seen in \cite{Noack}.

Our work is organized as follows. In the next section, we set up the equations of motion in the $3+1$-dimensional AdS black hole background. In section III, we explore the properties of the striped holographic superconductor system below and above the critical temperature. Then in section IV we extend our discussion to the insulator/striped superconductor phase transition. We conclude our results in the last section.

\section{The basic equations of motion in the four-dimensional AdS black hole background}
We consider the four-dimensional Einstein gravity coupled to a $U(1)$ gauge field with field strength $F_{\mu\nu}$ and a charged complex scalar field $\Psi$. The action reads
\begin{equation}\label{action}
S=\int d^4x\sqrt{-g}\Big[\frac{1}{16\pi G_N}(R-2 \Lambda)-\frac{1}{4}F^{\mu \nu}F_{\mu
\nu}-|\nabla \Psi-iqA \Psi|^2-V(|\Psi|)\Big].
\end{equation}
where the cosmological constant $\Lambda=-3/L^2$ with $L$ the $AdS$ radius and the scalar potential $V(|\Psi|)=m^2|\Psi|^2$ . From the action, we can derive field equations of motion as follows:
\begin{itemize}
\item
the equation of motion for the scalar field
\begin{equation}\label{eomPsi}
- \frac{1}{\sqrt{-g}}D_\mu\left(\sqrt{-g}g^{\mu\nu}D_\nu\Psi\right) + \frac{1}{2} \frac{\Psi}{|\Psi|} V'(|\Psi|) = 0
\end{equation}
where $D_\mu = \partial_\mu - iqA_\mu$,
\item
the equation for the Maxwell field
\begin{equation}\label{eomA}
\frac{1}{\sqrt{-g}}\partial_\mu\left(\sqrt{-g} F^{\mu\nu}\right) = i g^{\mu\nu} \left[\Psi^* D_\mu\Psi - \Psi (D_\mu\Psi)^* \right],
\end{equation}
\item
and the Einstein's equations
\begin{equation}\label{eomEinst}
R_{\mu\nu} - \frac12g_{\mu\nu} R - \frac{3}{L^2}g_{\mu\nu} =8\pi G_N T_{\mu\nu}
\end{equation}
\end{itemize}
where $T_{\mu\nu}= F_{\mu \rho} F_\nu{}^\rho - \frac{1}{4}g_{\mu\nu} F^{\rho\sigma} F_{\rho\sigma}
 - g_{\mu\nu} (D_\rho\Psi)(D^\rho\Psi)^* + \left[D_\mu\Psi (D_\nu\Psi)^* + D_\nu\Psi (D_\mu\Psi)^* \right]- g_{\mu\nu} V(|\Psi|)\vphantom{\frac14}$ is the energy-momentum tensor of the matter fields.

According to the AdS/CFT duality, this gravitational system at the classical level is required to study the strong coupling regime of the superconductor. We are interested in finding solutions to the classical equations of motion above,  whose boundary values are related to the parameters of the superconductor.  In this work, we will concentrate on the probe limit and neglect the backreaction of the gauge filed and the scalar field on the gravitational background.

\section{The inhomogeneous solutions}
In the probe limit, the Einstein equations admit the planar Schwarzschild AdS black hole solution
\begin{equation}\label{schwarMetric}
ds^2=L^2\big[-\frac{h(z)}{z^2}dt^2+\frac{1}{z^2h(z)}dz^2+\frac{1}{z^{2}}(dx^2+dy^2)\big],~~~{\rm with}~~~h(z)=1-\frac{z^3}{z_0^3}.
\end{equation}
Here, $z=z_0$ is the position of the horizon satisfying $h(z_0)=0$ while the AdS boundary locates at $z=0$. The Hawking temperature reads $T=\frac{3}{4\pi z_0 }$
which will be considered as the temperature of the dual gauge theory at the boundary. For simplicity, we will set $z_0=L=1$ in the following discussion.

In order to discuss the inhomogeneity, we will take the ansatz of the Maxwell field as $A_\mu=A_t(z,x)dt$ and set the scalar field $\Psi= \psi(z,x)$ to be real.  $A_t(z,x)$ and $\psi(z,x)$ are only two real fields we need to take into account. In the homogenous case limit,  the fields only depend on the radial coordinate $z$.

\subsection{Below the critical temperature}

Now we summarize the main results on the influence of the inhomogeneity in the superconductor observed in \cite{Pajer}.

Fixing the gauge with only non-zero electrostatic scalar potential, in the background (\ref{schwarMetric}),  the field equations of motion (\ref{eomPsi}) and (\ref{eomA}) can be rewritten as
\begin{eqnarray}\label{EOMS}
\psi^{''}+(\frac{h'}{h}-\frac{2}{z})\psi^{'}+\frac{\partial_x^2\psi}{h}+(\frac{A_t^2}{h^2}-\frac{m^2}{hz^2})\psi=0,\\
A_t^{''}+\frac{\partial_x^2 A_t}{h}-\frac{2\psi^2}{hz^2}A_t=0,
\end{eqnarray}
where the prime denotes the derivative with respect to $z$.

To study the effect of inhomogeneity, we are going to consider the following eletrostatic potential
 \cite{Pajer}, $A_t(z,x)=A_{t_{0}}(z)+A_{t_{1}}(z)Cos(Qx)$ with $Q$ the $x$-frequency of the mode or the spacial modulation.  $Q=0$ corresponds to the limiting homogeneous situation.  The
inhomogeneous term is introduced to source the charge density wave \cite{cdw}.
We expand the scalar field $\psi$ into Fourier space and look for the solution of the form i.e.,
\begin{eqnarray}\label{FieldFourier}
\psi(z,x)&=&\sum_{n=0}^{\infty}\psi_n(z)Cos(nQx).
\end{eqnarray}
In \cite{Pajer}, it was noted that higher modes are strongly suppressed in the Schwarzschild-AdS$_4$ spacetime. In principle, the higher modes can be included, but these make
the coupled equation for those modes increases.
Since we are only interested in the case of large modulation, we shall concentrate on the case with only two Fourier modes. Then substituting (\ref{FieldFourier}) into the equation (\ref{EOMS}), we obtain a series  of equations
\begin{eqnarray}\label{EomMode}
\psi_0^{''}+\big(\frac{h'}{h}-\frac{2}{z}\big)\psi_0^{'}+\big(\frac{A_{t_{0}}^2}{h^2}+\frac{A_{t_{1}}^2}{2h^2}-
\frac{m^2}{hz^2}\big)\psi_0+\frac{A_{t_{0}}A_{t_{1}}}{h^2}\psi_1=0, \nonumber \\
\psi_1^{''}+\big(\frac{h'}{h}-\frac{2}{z}\big)\psi_1^{'}+\big(\frac{A_{t_{0}}^2}{h^2}+\frac{3A_{t_{1}}^2}{4h^2}-
\frac{m^2}{hz^2}-\frac{Q^2}{h}\big)\psi_1+\frac{2A_{t_{0}}A_{t_{1}}}{h^2}\psi_0=0, \nonumber \\
A_{t_{0}}^{''}-\frac{2\psi_0^2+\psi_1^2}{hz^2}A_{t_{0}}-\frac{2\psi_0\psi_1}{hz^2}A_{t_{1}}=0, \nonumber \\
A_{t_{1}}^{''}-\big(\frac{2\psi_0^2+\frac{3}{2}\psi_1^2}{hz^2}+\frac{Q^2}{h}\big)A_{t_{1}}-\frac{4\psi_0\psi_1}{hz^2}A_{t_{0}}=0.
\end{eqnarray}

Near the AdS boundary $z\rightarrow0$, the behaviors of the scalar fields are related to the observables in the dual gauge theory as
\begin{eqnarray}\label{bdy2}
\psi(z,x)&=& z^{\vartriangle_-}(\psi_0^{(1)}+\psi_1^{(1)}CosQx)+ z^{\vartriangle_+}(\psi_0^{(2)}+\psi_1^{(2)}CosQx),\nonumber\\
A_t(z,x)&=&\mu(x)-\rho(x) z+\mathcal{O}(z),
\end{eqnarray}
where $\vartriangle_\pm=\frac{3}{2}\pm
\frac{1}{2}\sqrt{9+4 m^2}$. We will focus on $m^2=-2$ in this work. The  coefficients
$\mu(x)$ and $\rho(x)$ are the chemical potential and
the density of the charge in the dual field
theory. According to the
AdS/CFT dictionary, $\psi_{n}^{(1)}$ and $\psi_{n}^{(2)}$ for $n=0,1$ are normalizable, and one of them can be
the source while the other  be the response for
the dual operator $\psi_{n}^{(i)}\sim\mathcal{O}_{n}^{i}(i=1,2)$. The scalar operator $\mathcal{O}$ coupled to the scalar field $\psi(z,x)$ has the vacuum expectation value
\begin{eqnarray}\label{vev}
\langle\mathcal{O}^i\rangle=\psi_0^{(i)}+\psi_1^{(i)}CosQx,~~~i=1,2.
\end{eqnarray}

We can numerically integrate the equation group (\ref{EomMode}) from the horizon $z=1$ to the boundary $z=0$ by appropriately setting the boundary conditions. At the infinity, $ A_{t_{0}}(0)=\tilde{\mu}(1-\delta)$ and $ A_{t_{1}}(0)=\tilde{\mu}\delta$, so that we keep  $A_t(0,x)=\mu(x)=\tilde{\mu}[1-\delta(1-Cos Qx)]$ in (\ref{bdy2}). Here $\delta$ controls the strength of the inhomogeneity, where $\delta=1$ the system is completely inhomogeneous. Near the horizon, we expand the equations as a power series and impose $A_{t_{0}}(1)=A_{t_{1}}(1)=0$ for regularity. Then we can match our solution to the boundary conditions  $\psi_n(0)=0$ and $\delta A_{t_{0}}(0)=(1-\delta)A_{t_{1}}(0)$.
\begin{figure}
\centering
\includegraphics[width=.4\textwidth]{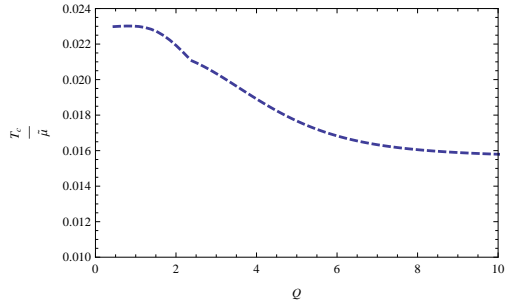}\hspace{0.1cm}
\caption{The critical temperature as a function of the spacial modulation $Q$ with the inhomogeneity parameter $\delta=0.4$.}
\label{figTc-Q}
\end{figure}


The spacial coordinate $x$ does not appear in  (\ref{EomMode}). The influence by the inhomogeneity is reflected through the spacial modulation $Q$ and the parameter $\delta$. The critical temperature does not depend on the $x$-position, but it is affected by the inhomogeneity through the modulation parameter $Q$. In figure \ref{figTc-Q}, we show the dependence of the critical temperature on  $Q$. We see that as $Q$ increases, the critical temperature  becomes smaller. The influence of $Q$ on the critical temperature is consistent with that  observed in\cite{Pajer,Siopsis1}.

\begin{figure}
\centering
\includegraphics[width=.4\textwidth]{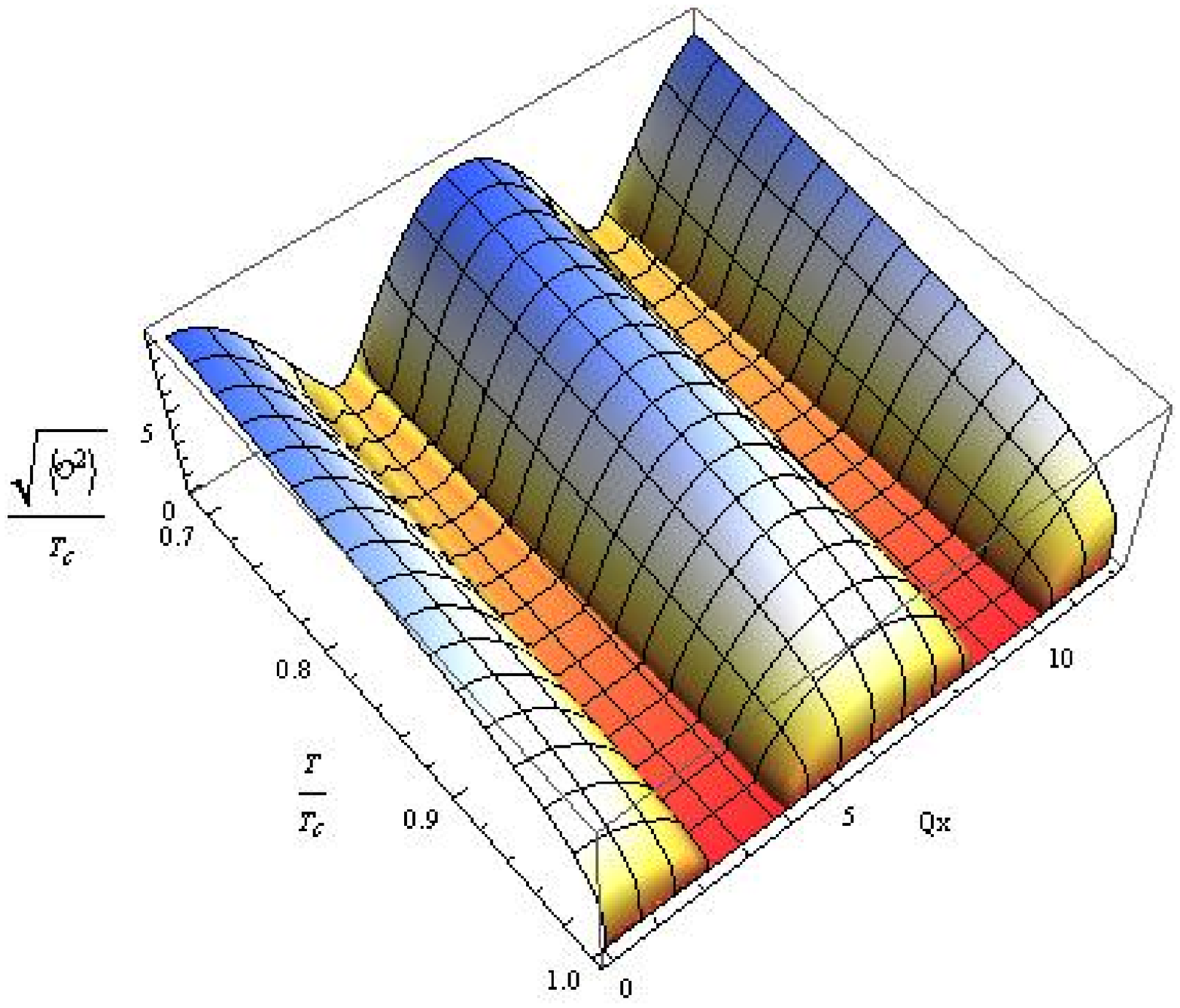}\hspace{0.6cm}
\includegraphics[width=.4\textwidth]{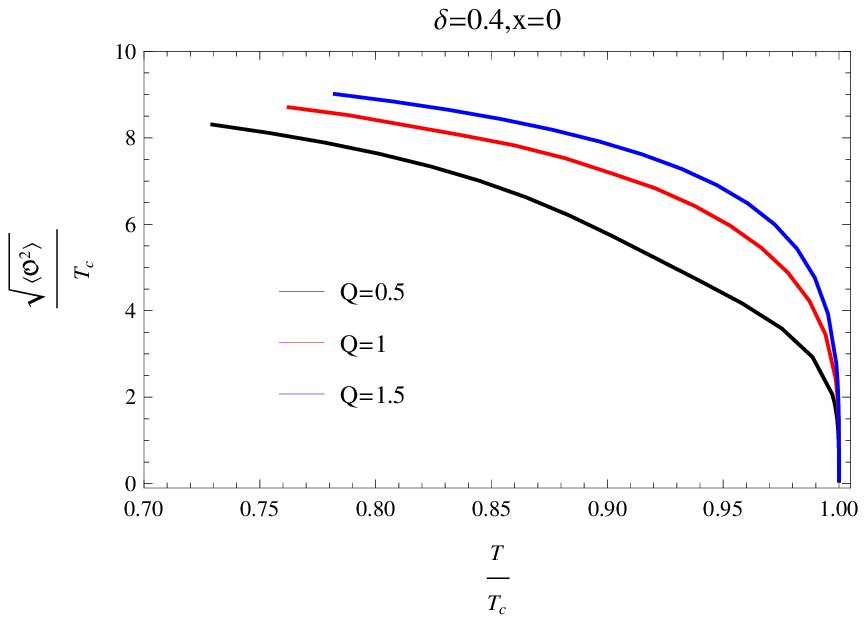}\hspace{0.1cm}
\caption{Left: The condensation with $Q=0.25$ and $\delta=0.4$. Right: The condensation at $x=0$ slice for some specified $Q$ with $\delta=0.4$.}
\label{figconden}
\end{figure}

Below the critical temperature, the scalar field starts to condensate near the horizon.
In figure \ref{figconden}, we show the behavior of the condensation. We see that when $Q$ is small there exists sharp superconducting  stripes  separated by stripes of normal phase. This agrees with the observation in \cite{Pajer,Siopsis1}.  When $Q$ becomes bigger, we also observe that the system is stiffer in the $x$ direction and the stripes are less pronounced.  Taking $x=0$, we show the dependence of the condensation gap on the parameter $Q$ in figure \ref{figconden}.  For choosing the same value of the scalar mass and $\delta$, qualitative features occur as we vary $Q$ in the condensations. As $Q$ increases, the condensation gap becomes higher, which means that the scalar hair can be difficult to be formed in the more inhomogeneous situation. This property holds at other values of $x$ as well.  This is in agreement of the dependence of $T_c$ on $Q$, showing that the inhomogeneity makes the condensation more difficult.

\subsection{Above the critical temperature}
\begin{figure}
\centering
\includegraphics[width=.4\textwidth]{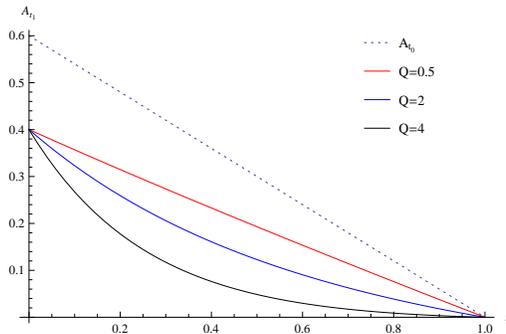}\hspace{0.2cm}
\caption{Profile of $A_{t_{0}}$ and $A_{t_{1}}$ with $\tilde{\mu}=1$. Here we set $\delta=0.4$ and change $Q$.}
\label{figAt1-z}
\end{figure}

Above the critical temperature, the scalar field vanishes $\psi(z,x)=0$ and thus the Maxwell equations in (\ref{EomMode}) reduce  into
\begin{eqnarray}\label{EomMaxwellMode}
A_{t_{0}}^{''}=0, \nonumber \\
A_{t_{1}}^{''}-\frac{Q^2}{h}A_{t_{1}}=0.
\end{eqnarray}
By imposing $A_t(z=0,x)=\tilde{\mu}[(1-\delta)+\delta Cos(Qx)]$ as before, we obtain $A_{t_{0}}=\tilde{\mu}(1-\delta)(1-z)$. $A_{t_{1}}$ can be got from numerically solving (\ref{EomMaxwellMode}) with boundary conditions $A_{t_{1}}(z=1)=0$ and $A_{t_{1}}(z=0)=\tilde{\mu}\delta$. Some sample solutions of $A_{t_{0}}$ and $A_{t_{1}}$ for different $Q$ are shown in figure \ref{figAt1-z}, where we have set $\tilde{\mu}=1$ without loss of generality. We see that the homogenous part $A_{t_{0}}$ is not affected by the spacial modulation while $A_{t_{1}}$ is suppressed by larger $Q$.

Once we have the gravitational background and the Maxwell field behavior, we are in a position to examine the dynamical behavior of the scalar field with the drop of the temperature. This can help us understand more on how the scalar condensation forms. In the inhomogeneous system, we expect that the study of the dynamical property of the scalar field with the decrease of the temperature can disclose the role of the inhomogeneity in the process of the condensation.

We will concentrate on the dynamical pair susceptibility first. In \cite{K. Maeda,YQ Liu-1}, it was found that the susceptibility can be an effective tool to probe the holographic superconductivity. In condensed matter physics, the dynamical pair susceptibility can be measured directly via the second order Josephson effect and it is believed that this quantity can give direct view on the origin of the superconductivity \cite{She}. In \cite{kuang-prd} it was found that the susceptibility can be a clear probe to identify the order of the phase transition in the holographic condensation. Here we expect to argue that the susceptibility is an effective probe to disclose the inhomogeneity in the striped condensation.

In the dictionary of the AdS/CFT correspondence, the dynamical susceptibility in the boundary field theory can be calculated from the dynamics of the fluctuations of the corresponding scalar field in the bulk AdS background in the gravity side. We can expand the scalar perturbation
as
$\Psi=\psi(z,x)e^{-i\omega t}$ where $\psi(z,x)$ has the form in
(\ref{FieldFourier}). Then equations of motion for the first two modes of the scalar field read as
\begin{eqnarray}\label{EomModepertuba}
\psi_0^{''}+\big(\frac{h'}{h}-\frac{2}{z}\big)\psi_0^{'}+\big[\frac{(\omega+ A_{t_{0}})^2}{h^2}+\frac{A_{t_{1}}^2}{2h^2}-
\frac{m^2}{hz^2}\big]\psi_0+\frac{A_{t_{1}}(\omega+ A_{t_{0}})}{h^2}\psi_1=0, \nonumber \\
\psi_1^{''}+\big(\frac{h'}{h}-\frac{2}{z}\big)\psi_1^{'}+\big[\frac{(\omega+ A_{t_{0}})^2}{h^2}+\frac{3A_{t_{1}}^2}{4h^2}-
\frac{m^2}{hz^2}-\frac{Q^2}{h}\big]\psi_1+\frac{2A_{t_{1}}(\omega+ A_{t_{0}})}{h^2}\psi_0=0,
\end{eqnarray}
where $A_{t_{0}}$ and $A_{t_{1}}$ satisfy equations (\ref{EomMaxwellMode}). The above equations
go back to the first two equations of (\ref{EomMode}) when $\omega=0$. We can solve equations (\ref{EomModepertuba}) by considering the infalling boundary conditions $\psi_I=(1-z)^{\frac{-i\omega}{4 \pi T}}[1+\tilde{\psi_I}^{(1)}(1-z)+\tilde{\psi_I}^{(2)}(1-z)^2+\cdots](I=0,1)$ near the horizon. Near the boundary, the scalar modes $\psi_I$ have the behaviors as $\psi_I\sim z\psi_I^{(1)}+z^2\psi_I^{(2)}$. The dynamical pair susceptibility can be identified by\cite{Son,KW}
\begin{equation}\label{chi1}
\chi=G^{R}=\frac{\psi^{(2)}}{\psi^{(1)}}=\frac{\psi_0^{(2)}+\psi_1^{(2)}CosQx}{\psi_0^{(1)}+\psi_1^{(1)}CosQx}
\end{equation}
by choosing $\psi^{(1)}$ as the source and $\psi^{(2)}$ as the response. In the condensed matter physics, the imaginary part of this quantity can be measured via second order Josephson effect and is proportional to the current through a tunneling junction \cite{She}.
\begin{figure}
\centering
\includegraphics[width=.4\textwidth]{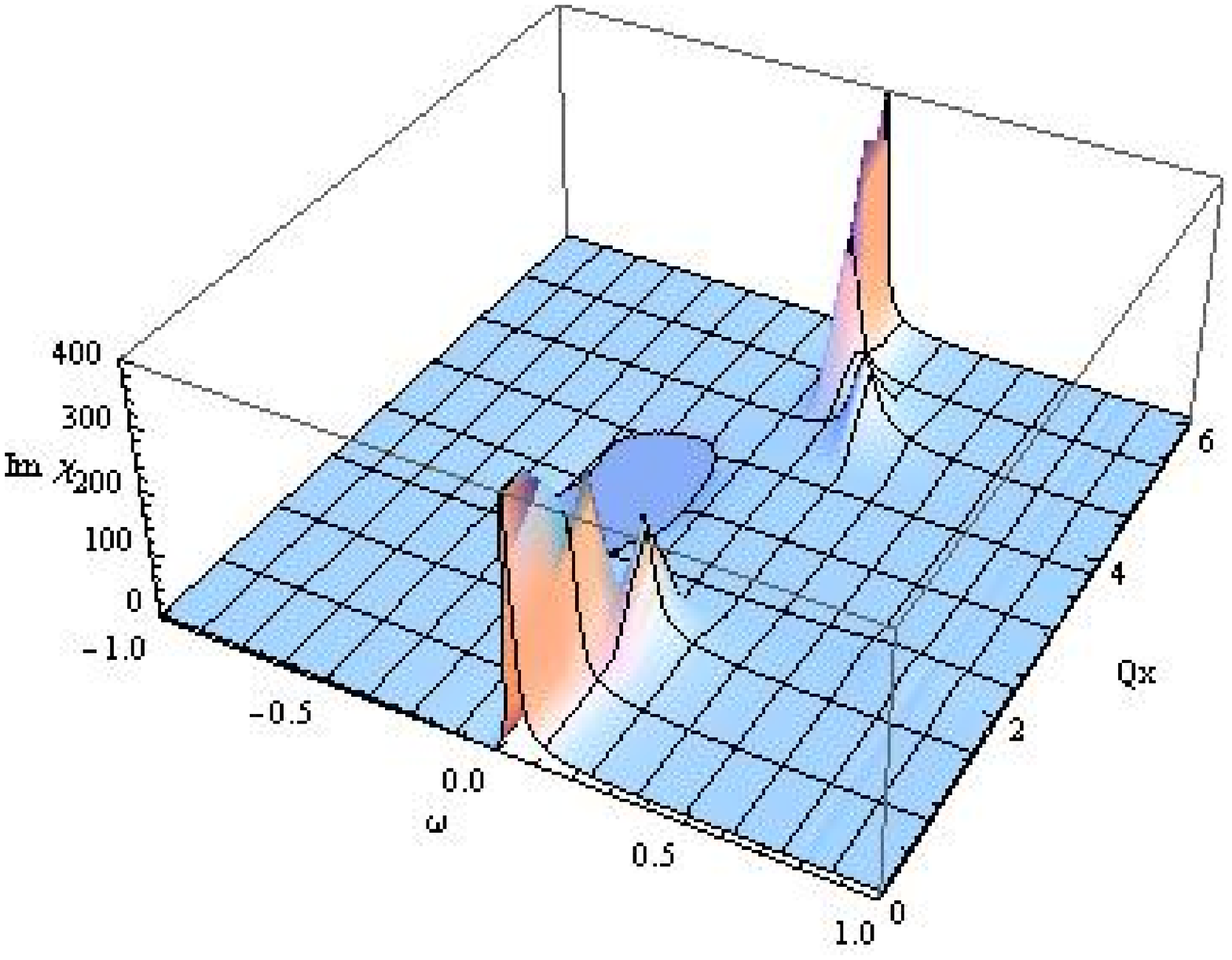}\hspace{0.2cm}
\includegraphics[width=.4\textwidth]{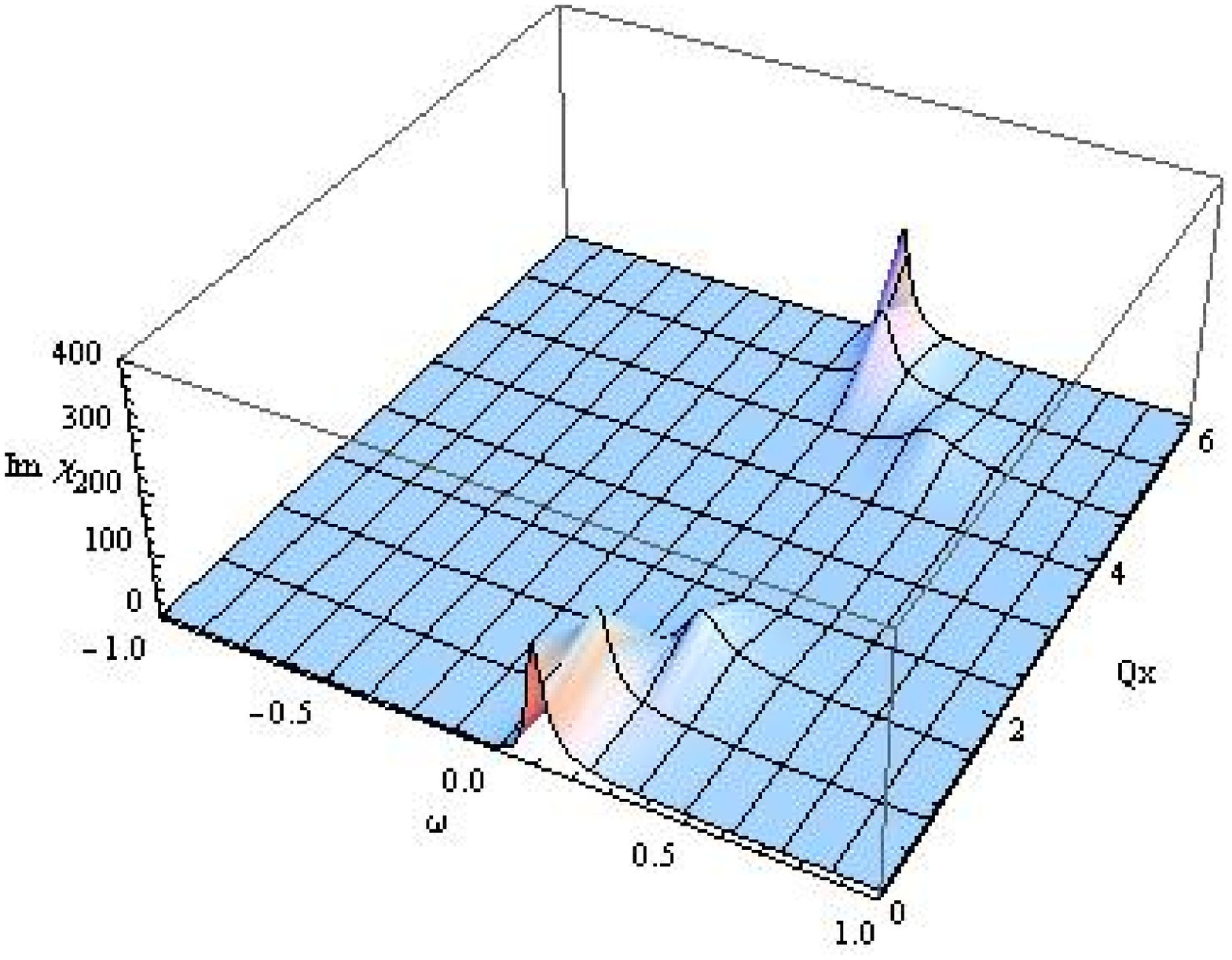}\hspace{0.2cm}
\includegraphics[width=.4\textwidth]{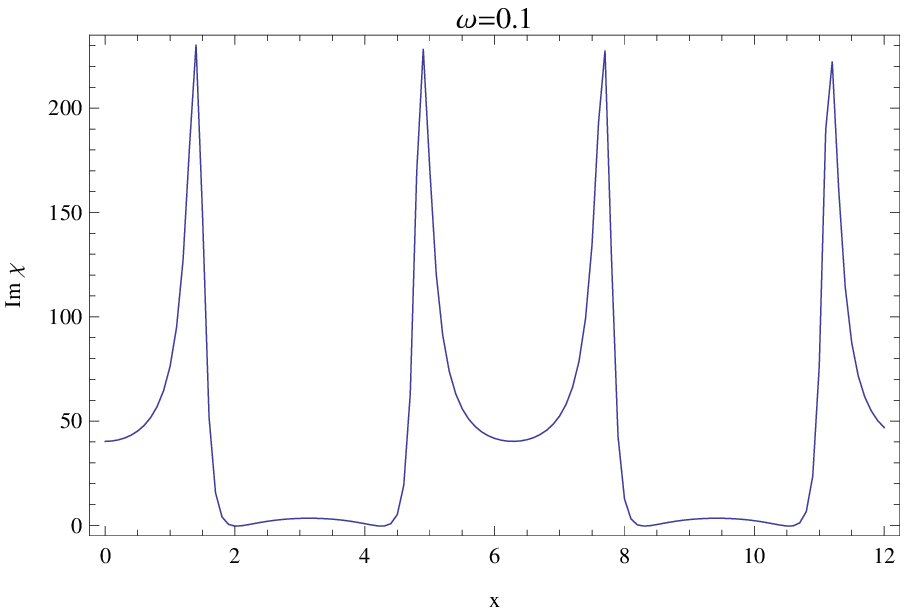}\hspace{0.2cm}
\includegraphics[width=.4\textwidth]{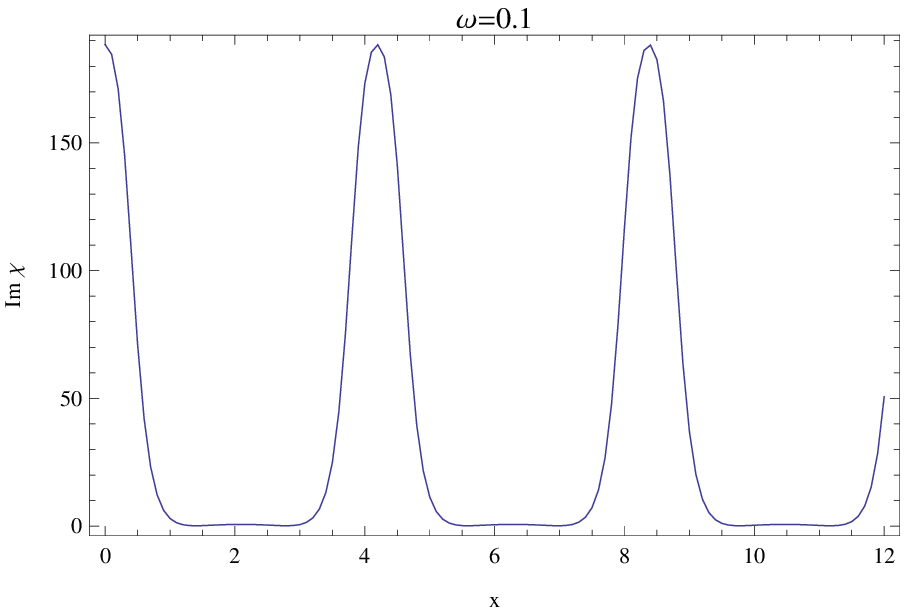}
\caption{The behavior of the imaginary part of pair susceptibility with $\delta=0.4$ at $T=1.1T_c$. The plots in the left panel are for $Q=1$ while in the right panel, the plots are for $Q=1.5$.}
\label{figwx-Imchi}
\end{figure}
\begin{figure}
\centering
\includegraphics[width=.3\textwidth]{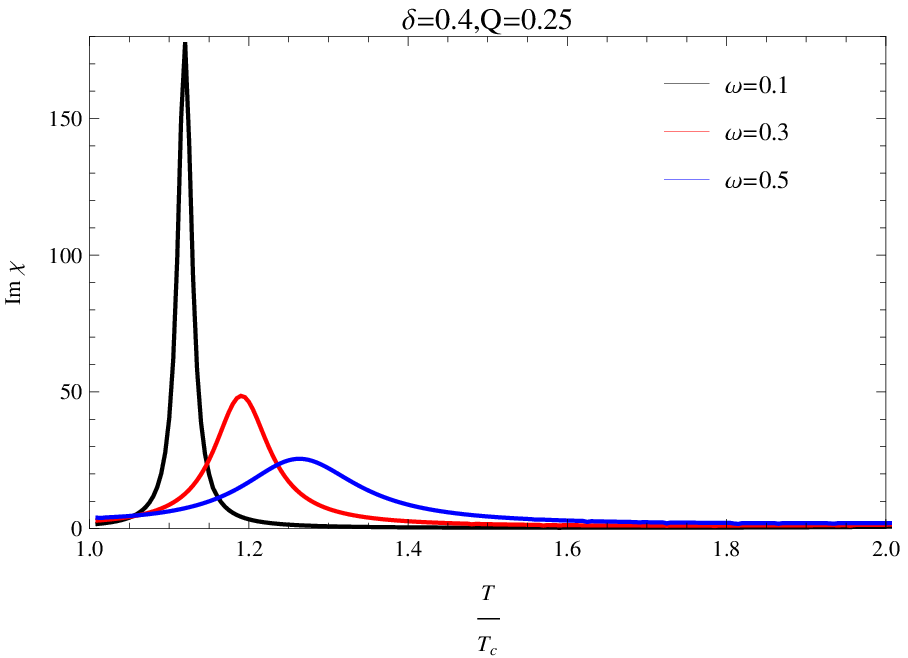}\hspace{0.3cm}
\includegraphics[width=.3\textwidth]{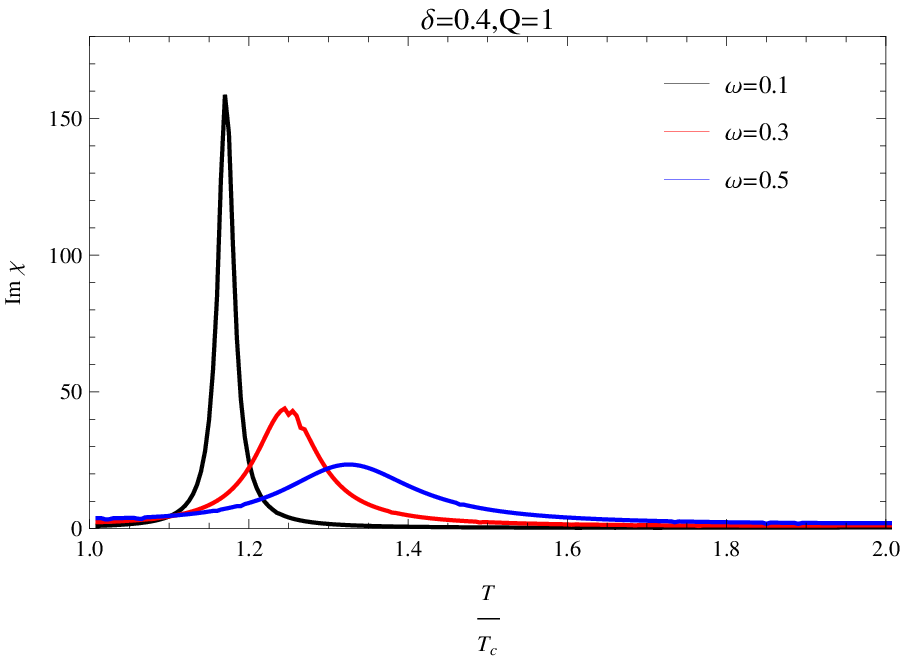}
\caption{The behavior of the imaginary part of pair susceptibility as the temperature approaches the critical temperature with different $Q$ and $x=0$.}
\label{figT-Imchi}
\end{figure}

We show our numerical results of the imaginary part of the dynamical pair susceptibility $\chi''$ at the fixed temperature $T=1.1T_c$ in figure \ref{figwx-Imchi}.  We observe that the peak of the imaginary part of the dynamical pair susceptibility becomes sharper and stronger for smaller $Q$. This actually explains that smaller $Q$ makes the phase transition easier to occur.  Besides due to the inhomogeneity, $\chi''$ is characterized by periodicity in the $x$ direction. The period increases as the wave vector $Q$ is deceased. This is reasonable because $Qx$ is always combined. The feature of period is more explicit in the 2D plots with a fixed frequency $\omega$ in the lower panel in figure \ref{figwx-Imchi}.

\begin{figure}
\centering
\includegraphics[width=.4\textwidth]{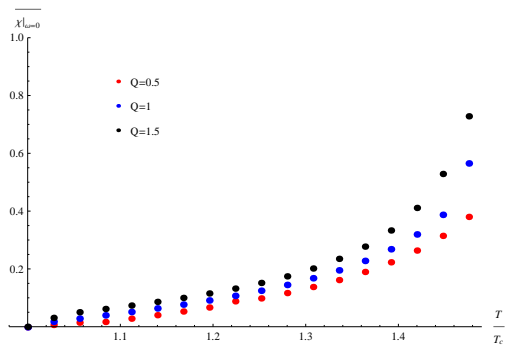}
\caption{The thermodynamic susceptibility as a function of the temperature with $\delta=0.4$ and samples of $Q$.}
\label{figT-1ToCHI}
\end{figure}

The behaviors of imaginary part of the dynamical pair susceptibility $\chi''$ for different $\omega$ when we approach the critical temperature from above  are shown in figure \ref{figT-Imchi}. We have fixed $\delta=0.4$ to see the influence of the spacial modulation.  For smaller frequency of the perturbed scalar field, the peak appears closer to the critical temperature and is higher and sharper. This implies that the scalar field with smaller energy will condensate easier near the horizon. While for the scalar field with higher energy, it is not easy to settle down to condensate.  With the change of the modulation parameter $Q$, we observe that for smaller $Q$, the peak of the imaginary part of the dynamical pair susceptibility is higher  for the same chosen frequency. The phenomenon is consistent with the observation in the behaviors of the critical temperature and the condensation that the larger spacial modulation makes the condensation harder.

Another phenomenological parameter we are interested is the thermodynamic susceptibility $\chi\mid_{\omega=0}$ which is the response function (\ref{chi1}) for the stationary source, i.e., $\omega=0$\cite{K. Maeda}.  It was argued that the static susceptibility can be an effective way to reflect the critical behavior near the condensation \cite{YQ Liu-1}.
At the critical temperature,  the thermodynamic susceptibility will be divergent at $T_c$. To detect this critical behavior, we calculate $\chi\mid_{\omega=0}$ from higher temperature to $T_c$. The results are shown in figure \ref{figT-1ToCHI}. We observe that the spacial modulation
does influence the tendency of the thermodynamic susceptibility when the critical point is approached. For the temperature of the system near the critical point,  the smaller spacial modulation has bigger thermodynamical susceptibility. This shows that for the smaller $Q$, the system is closer to the critical point so that the phase transition is easier to happen.

Besides the pair susceptibility, the conductivity was also found as an interesting parameter in studying holographic superconductor \cite{kuang-prd}.  In \cite{Pajer}, the property of conductivity and its relation to the modulation $Q$ has been studied. It was shown in their Fig.11 that when $Q$ is small, inhomogeneous conductivity becomes almost indistinguishable from the homogeneous one. When $Q$ grows, the conductivity deviates away from the homogeneous situation.  Thus both the conductivity and the pair susceptibility can be helpful to disclose the inhomogeneity in the holographic superconductor.

\section{Striped phases in the holographic insulator}
In the following we extend our discussion to the AdS soliton background.  The AdS soliton spacetime is described by
\begin{equation}\label{solitonMetric}
ds^2=L^2\big[\frac{1}{z^2h(z)}dz^2+\frac{1}{z^2}(-dt^2+dx^2)+\frac{h(z)}{z^{2}}d\eta^2\big],~~~~h(z)=1-\frac{z^3}{z_s^3}.
\end{equation}
There does not exist any horizon but a conical singularity at $z=z_s$, which can be removed by imposing a period $\beta=\frac{4\pi L^2 z_s}{3}$. Similarly, we will set $z_s=L=1$, $A_\mu=A_t(z,x)dt$ and $\Psi=\psi(z,x)$ to be real in this section.

In the background (\ref{solitonMetric}), we obtain the equations of motion for the scalar
field (\ref{eomPsi}) and gauge field (\ref{eomA}) in the forms
\begin{eqnarray}\label{soliton-EOMS}
\psi^{''}+(\frac{h'}{h}-\frac{2}{z})\psi^{'}+\frac{\partial_x^2\psi}{h}+(\frac{A_t^2}{h}-\frac{m^2}{hz^2})\psi=0, \nonumber\\
A_t^{''}+\frac{h'}{h}A_t^{'}+\frac{\partial_x^2 A_t}{h}-\frac{2\psi^2}{hz^2}A_t=0.
\end{eqnarray}
To study the inhomogeneous scalar condensation in the AdS soliton, we expand $A_t$ and $\psi$ into Fourier mode $A_t(z,x)=\sum_{n=0}^{\infty}A_{t_n}(z)Cos(nQx)$ and $\psi(z,x)=\sum_{n=0}^{\infty}\psi_n(z)Cos(nQx)$. Considering the modes to the first order, which is appropriate for not too small $Q$, we get the equations of motion below
\begin{eqnarray}\label{solitonEomMode}
\psi_0^{''}+\big(\frac{h'}{h}-\frac{2}{z}\big)\psi_0^{'}+\big(\frac{A_{t_{0}}^2}{h}+\frac{A_{t_{1}}^2}{2h}-
\frac{m^2}{hz^2}\big)\psi_0+\frac{A_{t_{0}}A_{t_{1}}}{h}\psi_1=0, \nonumber \\
\psi_1^{''}+\big(\frac{h'}{h}-\frac{2}{z}\big)\psi_1^{'}+\big(\frac{A_{t_{0}}^2}{h}+\frac{3A_{t_{1}}^2}{4h}-
\frac{m^2}{hz^2}-\frac{Q^2}{h}\big)\psi_1+\frac{2A_{t_{0}}A_{t_{1}}}{h}\psi_0=0, \nonumber \\
A_{t_{0}}^{''}+\frac{h'}{h}A_{t_{0}}^{'}-\frac{2\psi_0^2+\psi_1^2}{hz^2}A_{t_{0}}-\frac{2\psi_0\psi_1}{hz^2}A_{t_{1}}=0, \nonumber \\
A_{t_{1}}^{''}+\frac{h'}{h}A_{t_{1}}^{'}-\big(\frac{2\psi_0^2+\frac{3}{2}\psi_1^2}{hz^2}+\frac{Q^2}{h}\big)A_{t_{1}}-
\frac{4\psi_0\psi_1}{hz^2}A_{t_{0}}=0.
\end{eqnarray}
Before solving the equations of motion, we need to impose the boundary conditions at the tip $z=z_s$ and the AdS boundary $z=0$.

\begin{figure}
\centering
\includegraphics[width=.4\textwidth]{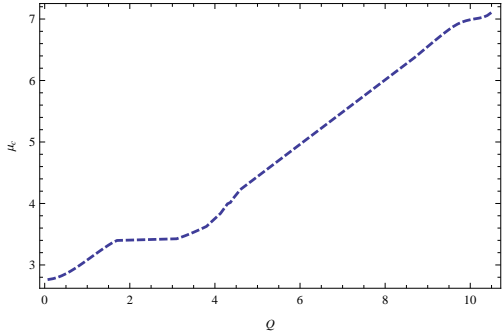}
\caption{The critical chemical potential as a function of the spacial modulation $Q$ with the inhomogeneity parameter $\delta=0.4$.}
\label{figMuc-Q}
\end{figure}

At the tip, the modes of the fields have asymptotic behavior
\begin{eqnarray}\label{soliton-tipbdy}
\psi_0&=&aa_0+bb_0(1-z)+cc_0(1-z)^2+\cdots, \nonumber \\
\psi_1&=&aa_1+bb_1(1-z)+cc_1(1-z)^2+\cdots, \nonumber \\
A_{t_{0}}&=&AA_0+BB_0(1-z)+CC_0(1-z)^2+\cdots, \nonumber \\
A_{t_{1}}&=&AA_1+BB_1(1-z)+CC_1(1-z)^2+\cdots.
\end{eqnarray}
We match the above behaviors with the boundary conditions near the infinite AdS boundary which are the same as those in the black hole background. We observe that when the chemical potential is stronger than a critical value $\mu_c$, the scalar field will condensate and a phase transition from the insulator to striped superconductor occurs. Similar to the phenomenon in black hole background, we find that larger spacial modulation makes the phase transition harder to occur. This can be seen from figure \ref{figMuc-Q} that $\mu_c$ increases as the spacial modulation $Q$ is enlarged.

\begin{figure}
\centering
\includegraphics[width=.4\textwidth]{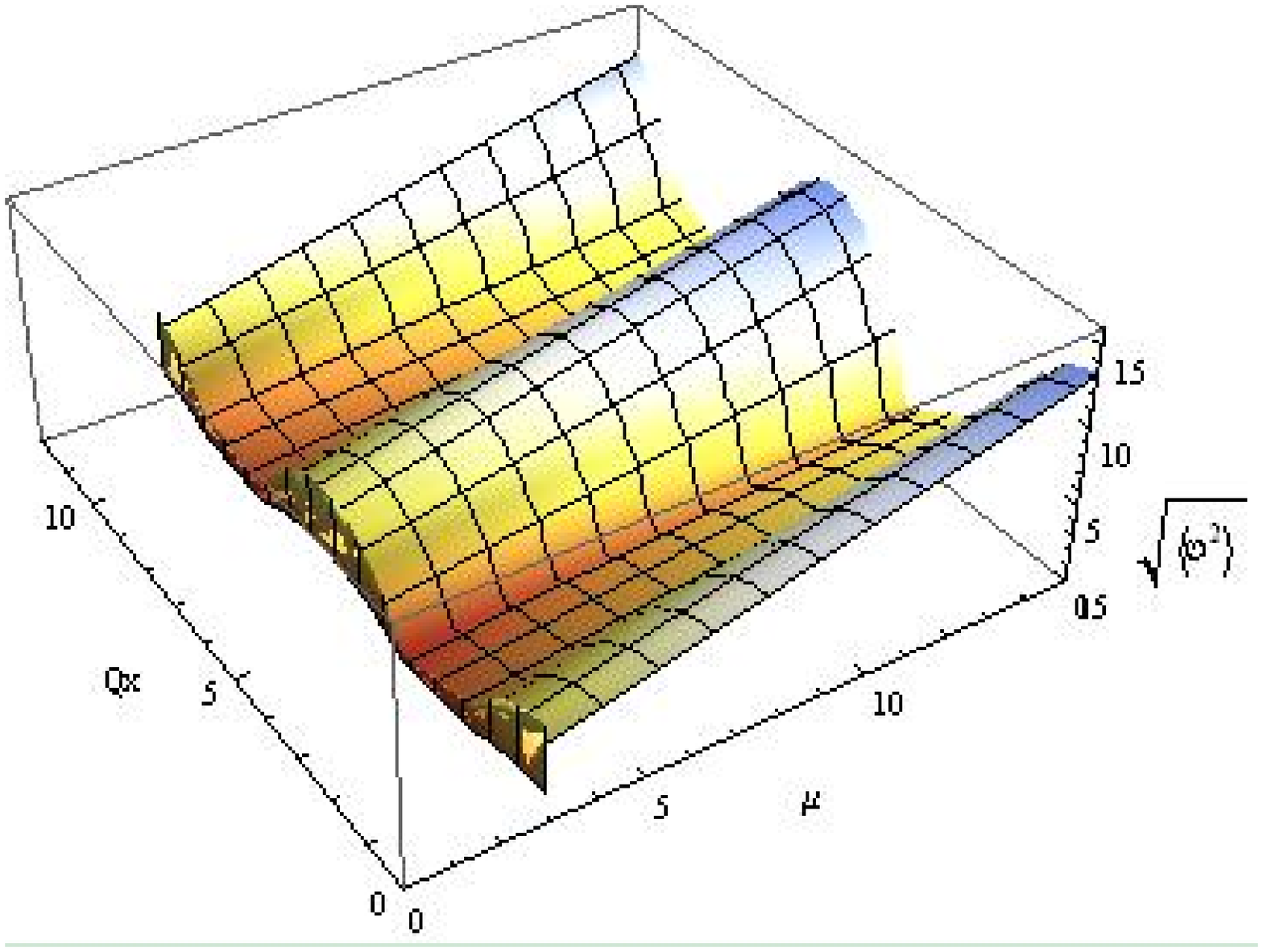}
\includegraphics[width=.4\textwidth]{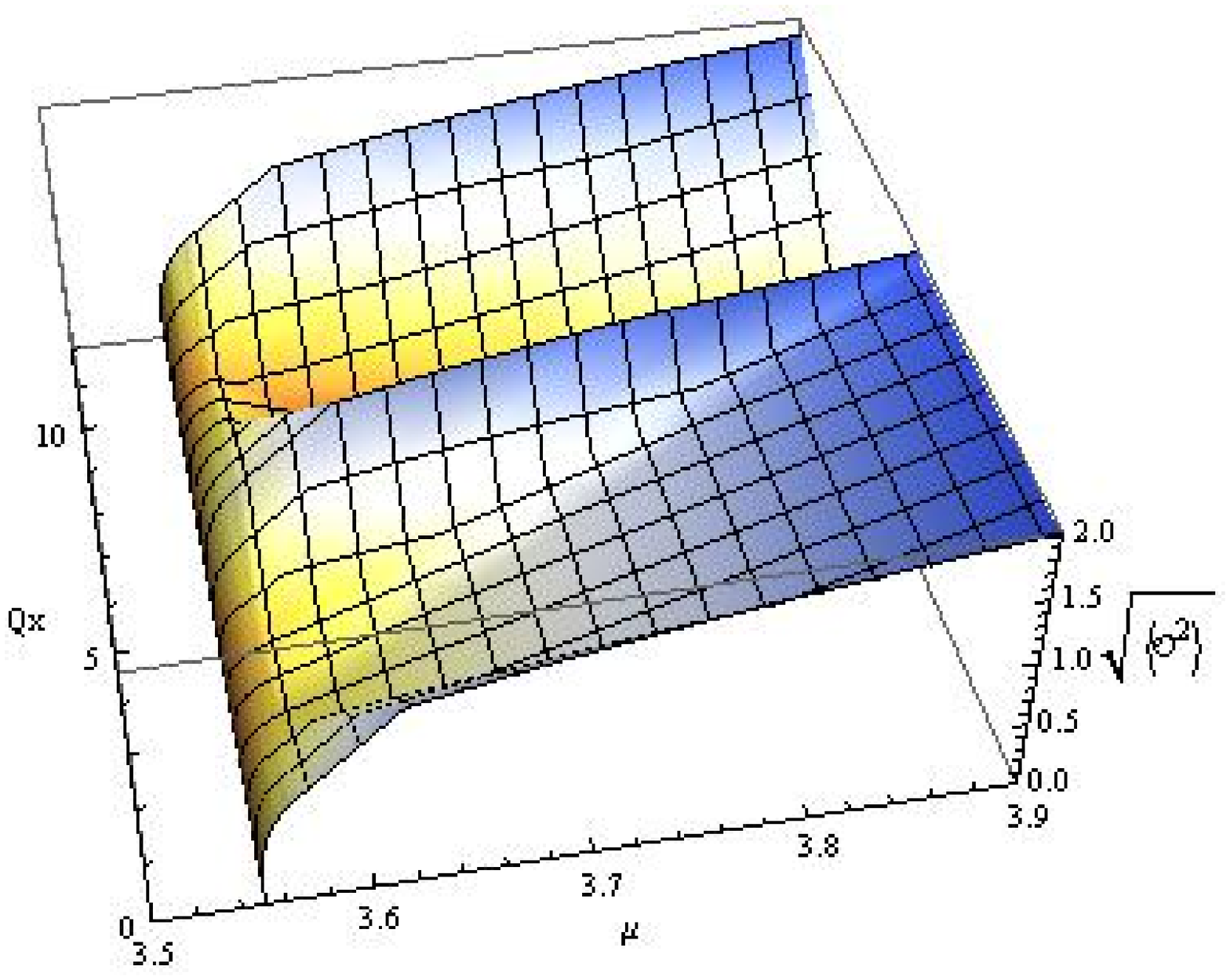}
\caption{3D condensation graph with $\delta=0.4$. Here we set $Q=0.5$(left) and $Q=2$(right).}
\label{fig3DsolConden}
\end{figure}

As $\mu$ is larger than the critical value, the vacuum expectation value $\langle\mathcal{O}^i\rangle=\psi_0^{(i)}+\psi_1^{(i)}CosQx,(i=1,2)$, where $\psi_0^{(i)}$ and $\psi_1^{(i)}$ can be read from the boundary values of $\psi_0$ and $\psi_1$, will continuously increase. We show the 3D condensation graph in figure \ref{fig3DsolConden}. It is clear that the stripes appear in the condensation.  This was also reported in \cite{Ge}. Here we find more behaviors in the condensation in the AdS soliton background.
For bigger $Q$, similar to that observed in the black hole case,  the system becomes stiffer in the $x$ direction so that the bulk condensation takes place everywhere at approximately the same $\mu$. The boundary system is in the superconducting phase everywhere, although the modulation is still visible.

\begin{figure}
\centering
\includegraphics[width=.4\textwidth]{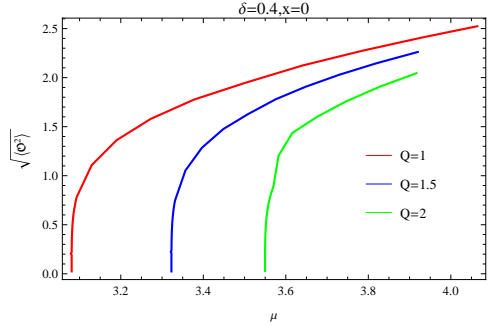}
\caption{The dependence of condensation on the chemical potential at $x=0$ for some specified $Q$.}
\label{fig2DsolConden}
\end{figure}

At $x=0$, the condensation behaviors for different $Q$ are shown in figure \ref{fig2DsolConden}. The critical chemical potentials are $3.0808$, $3.3227$ and $3.5496$ for different spacial modulation $Q=1, 1.5$ and $2$, respectively.
We also calculate the critical chemical potential $\mu_c\simeq1.7277$ for the  pure homogeneous background directly from equation (\ref{soliton-EOMS}). It is clear that
inhomogeneity enhances the value of the critical chemical potential.

\begin{figure}
\centering
\includegraphics[width=.4\textwidth]{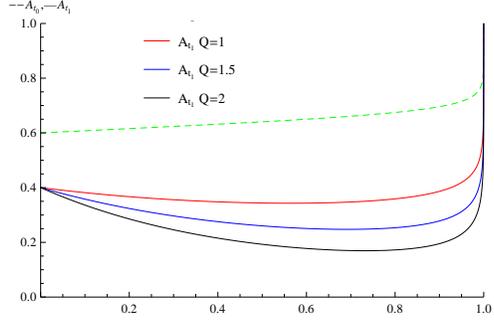}
\caption{The profile of the first two modes of Maxwell field with $\delta=0.4$.}
\label{figsolAt1z}
\end{figure}

Now we turn to study the dynamics in the AdS soliton and try to dig out the signature of the inhomogeneity in the scalar condensation in the AdS soliton background. We start from small  $\mu$ which is below  the critical chemical potential $\mu_c$, where there is no condensation, $\Psi=0$. Thus  the Maxwell equations  (\ref{solitonEomMode}) reduce into
\begin{eqnarray}\label{solitonEomModepsi0}
A_{t_{0}}^{''}+\frac{h'}{h}A_{t_{0}}^{'}=0, \nonumber \\
A_{t_{1}}^{''}+\frac{h'}{h}A_{t_{1}}^{'}-\frac{Q^2}{h}A_{t_{1}}=0.
\end{eqnarray}
Similar to the black hole case, we can assume the scalar perturbation in the form $\Psi=\psi(z,x)e^{-i\omega t}$. We can calculate the equations of motion for the first two modes of the scalar field
\begin{eqnarray}\label{solEomModepertuba}
\psi_0^{''}+\big(\frac{h'}{h}-\frac{2}{z}\big)\psi_0^{'}+\big[\frac{(\omega+ A_{t_{0}})^2}{h}+\frac{A_{t_{1}}^2}{2h}-
\frac{m^2}{hz^2}\big]\psi_0+\frac{A_{t_{1}}(\omega+ A_{t_{0}})}{h}\psi_1=0, \nonumber \\
\psi_1^{''}+\big(\frac{h'}{h}-\frac{2}{z}\big)\psi_1^{'}+\big[\frac{(\omega+ A_{t_{0}})^2}{h}+\frac{3A_{t_{1}}^2}{4h}-
\frac{m^2}{hz^2}-\frac{Q^2}{h}\big]\psi_1+\frac{2A_{t_{1}}(\omega+ A_{t_{0}})}{h}\psi_0=0,
\end{eqnarray}
Where $A_{t_{0}}$ and $A_{t_{1}}$ are the exact solutions of (\ref{solitonEomModepsi0}). We show several sample solutions of $A_{t_{0}}$ and $A_{t_{1}}$ with the change of the modulation $Q$   in figure \ref{figsolAt1z}. We set $\delta=0.4$ and $\tilde{\mu}=1$. Without loss of generality, we set the initial condition at the horizon $A_{t_{0}}(z=1)=A_{t_{1}}(z=1)=1$. We observe the similar behavior in the black hole case,  the spacial modulation $Q$ has no print on the zeroth mode $A_{t_{0}}$ but suppresses the first mode $A_{t_{1}}$ when $Q$ increases.
\begin{figure}
\centering
\includegraphics[width=.4\textwidth]{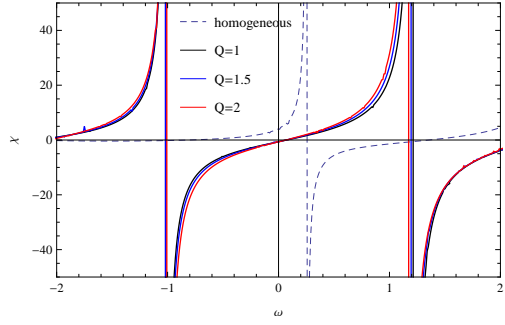}
\caption{The pair susceptibility in the soliton with $\mu_c/\mu=1.1$.}
\label{figsolw-chi}
\end{figure}

We try to get the pair susceptibility in the AdS soliton background. Solving the equations (\ref{solEomModepertuba}) with the boundary conditions $\psi_I=1+\tilde{\psi_I}^{(1)}(1-z)+\tilde{\psi_I}^{(2)}(1-z)^2+\cdots(I=0,1)$ at the tip, we find that in the AdS soliton
case, there is no imaginary part of the dynamical pair
susceptibility which is similar to that we observe in \cite{kuang-prd} .
This is because that both (\ref{solEomModepertuba}) and the boundary conditions are real. The vanishing of the imaginary part of $\chi$
implies that there is no dissipation in the response.
The real part of the pair susceptibility exists
as shown in figure \ref{figsolw-chi}.  The behavior of pair susceptibility in inhomogeneous soliton is similar to the shape in the homogeneous situation as shown in the dashed line in figure \ref{figsolw-chi}. When $Q$ is bigger, the pair susceptibility deviates more from that of the homogeneous result.

Besides the pair susceptibility, whether as we have seen in the black hole case, the conductivity can be the other probe to see the inhomogeneity in the AdS soliton background is a question to be asked.  Our solutions are inhomogeneous in the $x$-direction, while an electric field in the  $x$-direction sources  other independent perturbations even at linear order. The computation of $\sigma_x$ is therefore much more complicated, because we cannot simply introduce the electromagnetic perturbation $\delta A_x=A_x(z,x)e^{-i \omega t}$ as we did in the homogeneous case\cite{Q.Y. Pan-2}. We should consider all the possible components of Maxwell field, such as
\begin{eqnarray}\label{soliton-EMPertur}
z^2[h\partial_x^2{A_z}-\partial_x\partial_zA_x-h\partial_t^2A_z]=-2h[\psi\partial_z\chi-\chi\partial_z\psi-A_z(\psi^2+\chi^2)], \nonumber\\
z^2[h\partial_z^2{A_x}+h'\partial_zA_x-h\partial_z\partial_xA_z-h'\partial_xA_z-\partial_t^2A_x+\partial_t\partial_xA_t]=
-2[\psi\partial_x\chi-\chi\partial_x\psi-A_x(\psi^2+\chi^2)],\nonumber\\
z^2[-h\partial_z^2{A_t}-h'\partial_zA_t+h\partial_z\partial_tA_z+h'\partial_tA_z-\partial_x^2A_t+\partial_x\partial_tA_x]=
2[\psi\partial_t\chi-\chi\partial_t\psi-A_t(\psi^2+\chi^2)],
\end{eqnarray}
where the scalar field is $\Psi=\psi+i\chi$. Though the group of partial differential equations (\ref{soliton-EMPertur}) is difficult to solve, we can predict that the conductivity $\sigma_x$ is $x$ dependent. The spacial modulation $Q$ will qualitatively affect the conductivity since condensations of the the scalar field depends on $Q$.
 Thus the qualitative analysis shows that the conductivity along the direction of stripe can reflect the modulation factor $Q$ and can be distinguished from the homogeneous situation.
This is very different from the behavior of the conductivity perpendicular to the direction of the stripe. In \cite{Ge} it was  noticed that the conductivity perpendicular to the direction of the stripe is not much modified by the the spacial modulation, because as $\omega\rightarrow 0$ the conductivity becomes a delta function for various  spacial modes.  We hope to present the numerical result for  the conductivity along the direction of stripe in the future.

\section{conclusion}

We have studied the gravity duals of holographic superconductors in strongly interacting inhomogeneous systems. We have argued that the conductivity and the pair susceptibility, which are measurable quantities in the condensed matter physics, can be possible phenomenological indications to the inhomogeneity. Besides the AdS black hole background, we have also extended our discussion to the AdS soliton configuration. We have found that in the AdS soliton,  there is no imaginary part of the dynamical pair susceptibility. The  real part of the pair susceptibility can be used to distinguish the inhomogeneous case from the homogeneous situation. We argued that along the striped direction, the  conductivity can show the property of inhomogeneity. Further detailed exact understanding on the conductivity is called for.

\begin{acknowledgments}
We would like to thank Yunqi Liu for useful discussions. This work was supported by the National Natural Science Foundation of China.
\end{acknowledgments}

\end{document}